\documentclass[prc,aps,amssymb,twocolumn,showpacs,superscriptaddress]{revtex4}
\usepackage{graphicx}

\begin{document}

\title{$J/\psi$ transport in QGP and $p_t$ distribution at SPS and RHIC}

\author{Xianglei Zhu}
\affiliation{Physics Department, Tsinghua University, Beijing
100084, China}
\author{Pengfei Zhuang}
\affiliation{Physics Department, Tsinghua University, Beijing
100084, China}
\author{Nu Xu}
\affiliation{Nuclear Science Division,
   Lawrence Berkeley National Laboratory, Berkeley, California 94720, USA}
\date{\today}

\begin{abstract}
Combining the hydrodynamic equations for the QGP evolution and the
transport equation for the primordially produced $J/\psi$ in the
QGP, we investigate the $J/\psi$ transverse momentum distribution
as well as its suppression in the $\sqrt{s}$=17.3A GeV Pb-Pb
collisions at SPS and $\sqrt{s}$=200A GeV Au-Au collisions at
RHIC. The two sets of equations are connected by the $J/\psi$
anomalous suppression induced by its inelastic scattering with
gluons in the QGP. The calculated centrality dependence of
$J/\psi$ suppression and average transverse momentum square agree
well with the SPS data. From the comparison with the coalescence
model where charm quark is fully thermalized, our calculated
elliptical flow of the primordially produced $J/\psi$ is much
smaller. This may be helpful to differentiate the $J/\psi$
production mechanisms in relativistic heavy ion collisions.
\end{abstract}

\maketitle

\section{Introduction}
Many mechanisms are proposed to explain the phenomenon of
charmonium suppression\cite{matsui} in relativistic heavy ion
collisions\cite{NA50}. In models\cite{qiu, blaizot, xu1, wang,
polleri2, capella, xu2, urqmd, hsd}, the charmonium is supposed to
be created by the hard processes in the initial state, on the way
out, it firstly collides inelastically with spectators which leads
to the normal suppression\cite{vogt,gerschel}. The anomalous
suppression\cite{NA50} is attributed to different mechanisms:
collisions with spectators\cite{qiu}, instantly melting in
Quark-Gluon Plasma (QGP)\cite{blaizot}, collisions with gluons in
QGP\cite{xu1, wang, polleri2} or collisions with hadronic
comovers\cite{capella, xu2, urqmd, hsd}. Recently, there appeared
another kind of models\cite{braun, gorenstein} based on the
assumption that charmonium is created at QCD hadronization
according to statistical law, which is the extrapolation of the
well-tested thermal model\cite{thermal} for light hadrons. There
is also a two component model\cite{loic} which mixes these two
kinds of mechanisms. With some adjustable parameters in each
mechanism, almost all models can describe the $J/\psi$ anomalous
suppression data at SPS very well. In order to differentiate these
models, we need more information such as open charm, $\chi_c$ and
$\psi'$, and transverse momentum ($p_t$) distribution. In this
Letter, we concentrate on the observables related to the $p_t$
distribution of $J/\psi$, the average transverse momentum
$\left<p_t^2\right>$ and the anisotropic asymmetry parameter
($v_2$) of $J/\psi$ as well as the $p_t$ integrated anomalous
suppression in the heavy ion collisions at SPS and RHIC.

We follow the method in Refs. \cite{xu1, wang, polleri2} to treat
the anomalous suppression. We neglect the formation time and
assume that the charmonium is created instantaneously after the
binary collisions. On the way out, it collides firstly with
spectators and then with gluons in the QGP. The former and the
later are, respectively, the origin of normal and anomalous
suppression in our approach. The QGP evolution is calculated with
a 2 + 1 dimensional boost invariant relativistic
hydrodynamics\cite{bjorken}. Since $J/\psi$ is heavy enough, a
classical Boltzmann-type transport equation in the transverse
phase space\cite{zhuang} is used to describe the evolution of its
transverse distribution function. We assume that the local
equilibrium is reached at a proper time $\tau_0$ when the normal
suppression has ceased and the anomalous suppression starts.
Therefore, the initial condition of the transport equation is
determined by the normal suppression. We neglect the elastic
collisions between charmonium and particles in the medium, for the
much larger charmonium mass than the typical temperature of the
medium.

The paper is organized as follows. We describe the evolution of
the medium in Section 2, give the details of the cross sections
between charmonium and particles in Section 3, discuss the
transverse transport equation in Section 4, and show the numerical
results in Section 5. Finally we conclude in Section 6.

\section{Medium Evolution }

As in Ref.\cite{tau0}, we assume that the produced medium reaches
local equilibrium at a proper time $\tau_0 = 0.8$ fm/c for
$\sqrt{s} = 17.3$A GeV Pb-Pb collisions at SPS and $0.6$ fm/c for
$\sqrt{s} = 200$A GeV Au-Au collisions at RHIC. The consequent
evolution is described with relativistic hydrodynamics,
\begin{equation}
\partial_{\mu}T^{\mu\nu}=0, \ \ \ \partial_{\mu}N^{\mu}=0,
\end{equation}
where $T^{\mu\nu}=(\epsilon+p)u^{\mu}u^{\nu}-g^{\mu\nu}p$ is the
energy-momentum tensor and $N^{\mu}=nu^{\mu}$ the baryon current
with four-velocity $u^\mu$ of the fluid cell, energy density
$\epsilon$, pressure $p$ and baryon density $n$. In our
calculation, we use Bjorken's hydrodynamical model\cite{bjorken}.
With the Hubble-like longitudinal expansion and boost invariant
initial condition, the hydrodynamical quantities are functions of
the proper time $\tau =\sqrt {t^2-z^2}$ and transverse coordinates
only, and the equations can be simplified as
\begin{eqnarray}\label{hydro}
\partial_{\tau}E+\triangledown\cdot{\vec M} &=& -(E+p)/{\tau}\
,\nonumber\\
\partial_{\tau}M_x+\triangledown\cdot(M_x\vec{v}) &=& -M_x/{\tau}-\partial_xp\ ,
\nonumber\\
\partial_{\tau}M_y+\triangledown\cdot(M_y\vec{v}) &=& -M_y/{\tau}-\partial_yp \
,\nonumber\\
\partial_{\tau}R+\triangledown\cdot(R\vec{v}) &=& -R/{\tau}
\end{eqnarray}
with the definitions $E=(\epsilon+p){\gamma}^2-p$, ${\vec
M}=(\epsilon+p){\gamma}^2{\vec v}$ and $R=\gamma n$, where
$\gamma$ is the Lorentz factor.

To close the hydrodynamical equations we need to know the equation
of state of the medium. We follow Ref.\cite{sollfrank} where the
deconfined phase at high temperature is an ideal gas of massless
$u$, $d$ quarks, 150 MeV massed $s$ quarks and gluons, and the
hadron phase at low temperature is an ideal gas of all known
hadrons and resonences with mass up to 2 GeV\cite{pdg}. There is a
first order phase transition between these two phases. In the
mixed phase, the Maxwell construction is used. The mean field
repulsion parameter and the bag parameter are chosen as $K$=450
MeV fm$^3$\cite{sollfrank} and $B^{1/4}$=236 MeV to obtain the
critical temperature $T_c=165$ MeV at vanishing baryon number
density.

We assume that the medium maintains chemical and thermal
equilibrium until the energy density of the system drops to a
value of $60$ MeV/fm$^3$, when the hadrons decouple and their
momentum distributions are fixed. We have tested that a slight
modification on the freeze-out condition will not change our
conclusions.

At SPS energy, we follow Ref.\cite{hydroinit} to set the initial
energy density and baryon number density with the wounded nucleon
number density,
\begin{eqnarray}
n_{wn}({\vec b},{\vec x_t})&=&T_A({\vec
x_t})(1-e^{-\sigma_{NN}T_B({\vec x_t}-{\vec b})})
\nonumber\\
&+&T_B({\vec x_t}-{\vec b})(1-e^{-\sigma_{NN}T_A({\vec x_t})})\ ,
\end{eqnarray}
where $T_{A(B)}({\vec x_t}) =
\int_{-\infty}^{\infty}dz\,\rho_{A(B)}({\vec x_t}, z)$ is the
thickness function of nuclear A(B) with nuclear density profile
$\rho_{A(B)}$. We choose Woods-Saxon profile,
$\rho(r)=\frac{\rho_0}{e^{(r-R)/\xi}+1}$. The parameters $R$ and
$\xi$ for $^{208}Pb$ and $^{197}Au$ are from Ref.\cite{JVV}. At
RHIC energy, we follow Refs.\cite{kolbrhic1, kolbrhic2} and assign
75\% of the entropy density to soft contribution which is
proportional to the wounded nucleon number density and 25\% to
hard contribution which is proportional to the binary collision
number density $n_{bc}({\vec b},{\vec x_t})=T_A({\vec
x_t})T_B({\vec x_t}-{\vec b})\sigma_{NN}$. The baryon density is
obtained by adjusting the entropy per baryon to
250\cite{kolbrhic2}. The nucleon-nucleon inelastic cross section
$\sigma_{NN}$ is 32 mb at SPS and increases to 41 mb at RHIC.

In peripheral collisions or in the peripheral region of central
collisions, the system can not reach thermalization even if the
energy density is larger than the critical value for phase
transition. In order to describe this effect, we incorporated a
cut in the initial condition. We assume that the medium will
hadronize and decouple at the initial time if the local entropy
density is less than a critical value $s_c$. This critical value
will be an adjustable parameter of our approach, and will be fixed
by fitting the $J/\psi$ anomalous suppression data at SPS. The
same value will be used in the calculation of the hydrodynamics at
RHIC energy.

We use the well tested RHLLE algorithm\cite{schneider, Rischke} to
solve the hydrodynamical equations numerically. The simple first
order operator splitting method is used to extrapolate the
original one dimensional RHLLE algorithm to two dimensions.

\section{Dissociation Cross-Sections}
In the QGP phase, we consider gluon dissociation process
($g+\Psi\rightarrow c+\bar c$) only. We will use $\Psi$ as the
shorthand notation of charmonium in this Letter. The
gluon-$J/\psi$ dissociation cross section can be obtained from the
perturbative calculation with non-relativistic and Coulomb
potential approximation for the $c\bar c$ system in the
vacuum\cite{peskin},
\begin{equation}\label{gluon}
\sigma_g^\psi(\omega)=A_0\frac{(\omega/\epsilon_{\psi}-1)^{3/2}}{(\omega/\epsilon_{\psi})^5}
\end{equation}
with $A_0=(2^{11}\pi/27)(m_c^3\epsilon_{\psi})^{-1/2}$, where
$\omega=p_{\Psi}^{\mu}p_{g\mu}/m_{\Psi}=(s-m_{\Psi}^2)/(2m_{\Psi})$
is the gluon energy in the rest frame of $\Psi$, $\epsilon_{\Psi}=
2m_D-m_{\Psi}$ the binding energy of $\Psi$, $m_c=m_D=1.87 $GeV
and $m_{\psi}=3.1 $GeV are, respectively, the charm quark mass and
$J/\psi$ mass. In a similar way, the gluon dissociation cross
section for $\psi'$\cite{oh} and $\chi_c$\cite{wang} are
\begin{eqnarray}
\label{psiprimeandchi}
\sigma_g^{\psi'}(\omega) &=& 16A_0\frac{(\omega/\epsilon_{\psi'}-1)^{3/2}
(\omega/\epsilon_{\psi'}-3)^2}{(\omega/\epsilon_{\psi'})^7}\
,\nonumber\\
\sigma_g^{\chi_c}(\omega) &=& 4A_0(\omega/\epsilon_{\chi_c}-1)^{1/2} \nonumber\\
&\times&\frac{\left(9(\omega/\epsilon_{\chi_c})^2-20(\omega/\epsilon_{\chi_c})+12\right)}
{(\omega/\epsilon_{\chi_c})^7},
\end{eqnarray}
The masses of $\psi'$ and $\chi_c$ are $3.7$ GeV and $3.5$ GeV,
respectively. The $J/\psi$ from the feed-down of $\psi'$ is about
10\% of the total final $J/\psi$'s in pp collisions. For
simplicity, we neglect the contribution of $\psi'$ to $J/\psi$ in
our calculation. Therefore, 40\% of the final state $J/\psi$'s are
from the feed-down of $\chi_c$ and others are created
directly\cite{wang}.

We have neglected the medium effect on the dissociation cross
sections. In relativistic heavy ion collisions, the medium
produced is quite finite and evolves very fast, perhaps the medium
effect is not so important as that in a static and infinite
medium.

In the hadron phase, the most populated hadrons are pions. There
are many effective models that can calculate the inelastic cross
sections between charmonium and hadrons\cite{barnes}. For
$J/\psi$, the dissociation cross section is about a few mb which
is comparable to the gluon dissociation cross section. If the QGP
is created in relativistic heavy ion collisions, the thermalized
hadron phase is at the later time of the evolution, and at that
time the number density of hadrons is very small compared with the
number density of gluons at the early time. Although the hadron
dissociation may be important to the consideration of $\chi_c$ and
$\psi'$ suppression, we neglect the contribution of hadrons in the
calculation for $J/\psi$.

\section{Charmonium Transport }
We define the full phase space distribution function of charmonium
at a global time t as $f_{\Psi}(\vec{x},\vec{p},t)$. The transport
equation describing the evolution of $f_{\Psi}$ is
\begin{equation}
\label{transport}
p^{\mu}\partial_{\mu}\,f_{\Psi}(\vec{x},\vec{p},t)=
-C^{\Psi}_{D}(\vec{x},\vec{p},t)\,f_{\Psi}(\vec{x},\vec{p},t)\ .
\end{equation}
The l.h.s is the drift term and the r.h.s is the dissociation term
due to the inelastic collisions between charmonium and
constituents of the medium. The coefficient $C_D^{\Psi}$ of the
dissociation term in Eq. (\ref{transport}) is
\begin{equation}\label{loss}
C^{\Psi}_{D}(\vec{x},\vec{p},t)=\frac{1}{2}\int
\frac{d^3\vec{k}}{(2\pi)^3
2E_k}\sigma^{\Psi}_{D}(s)4F_{g\Psi}f_{g}(\vec{x},\vec{k},t)
\end{equation}
with $E_k=\sqrt{m^2_{g}+k^2}$ and flux factor
$F_{g\Psi}=\sqrt{(p^{\mu}_{\Psi}p_{g\mu})^2-m^2_{\Psi}m^2_{g}}$,
where $f_{g}$ is the Lorentz invariant distribution function of
the constituents, and $\sigma^{\Psi}_{D}(s)$ is the dissociation
cross section for the charmonium in the medium. It is obvious that
$C^{\Psi}_{D}$ is Lorentz invariant.

Considering the boost invariant initial condition at the proper
time $\tau_0$, it is convenient to use momentum rapidity
$y=1/2\,\ln\left((E+p_z)/(E-p_z)\right)$ and space-time rapidity
$\eta=1/2\,\ln\left((t+z)/(t-z)\right)$ as longitudinal variables.
With the assumption that the charmonium is produced at the point
$t=0$ and $z=0$ due to the large Lorentz contraction of nucleus
and short duration time of the initial hard processes in high
energy heavy ion collisions, the charmonium momentum rapidity is
equal to its space-time rapidity, $y=\eta$. The distribution
function can then be rewritten as
\begin{equation}\label{dis}
f_{\Psi}(\vec{x_t},\eta,\vec{p_t},y,\tau)=
\frac{\tau_0}{\tau}f_{\Psi}(\vec{x_t},\vec{p_t},y,\tau)\delta(\eta-y)\
.
\end{equation}
After integration over $\eta$, the l.h.s. of the transport
equation (\ref{transport}) is simplified as
\begin{equation}\label{left}
\frac{\tau_0}{\tau}(m_t \frac{\partial}{\partial
\tau}+\vec{p_t}\cdot\frac{\partial}{\partial
\vec{x_t}})f_{\Psi}(\vec{x_t},\vec{p_t},y,\tau)\ ,
\end{equation}
and the r.h.s becomes
\begin{equation}\label{right}
-\frac{\tau_0}{\tau}C^{\Psi}_D
(\vec{x_t},y,\vec{p_t},y,\tau)f_{\Psi}(\vec{x_t},\vec{p_t},y,\tau)\
.
\end{equation}

In Bjorken's hydrodynamics, the longitudinal rapidity of medium
flow equals its space-time rapidity. Therefore, $\Psi$'s rapidity
is zero in the longitudinal rest frame of the medium flow. Due to
the Lorentz invariance of $C^{\Psi}_D$, we can calculate it in
this rest frame. Furthermore, we assume that the temperature and
baryon chemical potential in the rest frame is $\eta$ independent,
since the rapidity region of the detected $\Psi$'s is very narrow
and near the central rapidity region. With this assumption,
$C^{\Psi}_D$ becomes independent of the $\Psi$ rapidity, the
transport equation can then be simplified as
\begin{eqnarray}
(m_t \frac{\partial}{\partial
\tau}&+&\vec{p_t}\cdot\frac{\partial}{\partial
\vec{x_t}})f_{\Psi}(\vec{x_t},\vec{p_t},y,\tau)\nonumber\\
&=& -C^{\Psi}_D
(\vec{x_t},\vec{p_t},\tau)f_{\Psi}(\vec{x_t},\vec{p_t},y,\tau)\ .
\end{eqnarray}
After integration over the rapidity $y$, it is reduced to
\begin{eqnarray} \label{ptransport}
(m_t
\frac{\partial}{\partial
\tau}&+&\vec{p_t}\cdot\frac{\partial}{\partial
\vec{x_t}})f_{\Psi}(\vec{x_t},\vec{p_t},\tau)\nonumber\\
&=& -C^{\Psi}_D
(\vec{x_t},\vec{p_t},\tau)f_{\Psi}(\vec{x_t},\vec{p_t},\tau)\ .
\end{eqnarray}

With Cooper-Frye formula\cite{CFformula}, the number of $\Psi$'s
at proper time $\tau$ is
\begin{eqnarray}
\label{ntau}
N_{\Psi}(\tau)&=&\frac{1}{(2\pi)^3}\int\,d^2\vec{p_t}\,d^2\vec{x_t}\,dy\,d\eta\,\tau\,m_t\cosh(y-\eta)\nonumber\\
& &f_{\Psi}(\vec{x_t},\eta,\vec{p_t},y,\tau)\nonumber \\
&=&\frac{1}{(2\pi)^3}\int\,d^2\vec{p_t}\,d^2\vec{x_t}\,\tau_0\,m_t
f_{\Psi}(\vec{x_t},\vec{p_t},\tau)\ .
\end{eqnarray}
Since the normal suppression of $\Psi$ has ceased before the
starting time of the medium evolution, the number of $\Psi$'s at
$\tau_0$ is
\begin{equation}
\label{ntau0} N_\Psi(\tau_0)=\int
f_N(\vec{x_t},\vec{p_t}|\vec{b})\,d^2\vec{x_t}d^2\vec{p_t}\ ,
\end{equation}
where $f_N$ is the $\Psi$ transverse distribution function at
impact parameter $\vec b$ after the normal
suppression\cite{zhuang},
\begin{eqnarray}\label{initialf}
f_N(\vec{x_t},\vec{p_t}|\vec{b})&=&\frac{\sigma^{\Psi}_{NN}}{\pi}\int
dz_A dz_B
\rho_A(\vec{x_t},z_A)\rho_B(\vec{x_t}-\vec{b},z_B)\nonumber\\
&\times&e^{-\sigma_{abs}\left(T_A(\vec{x_t},z_A,+\infty)+T_B(\vec{x_t}-\vec{b},-\infty,z_B)
\right)}\nonumber\\
&\times&{1\over \left<p_t^2\right>}e^{-p_t^2/\left<p_t^2\right>}\
\end{eqnarray}
with averaged transverse momentum square\cite{huefner88}
\begin{eqnarray}
\label{pt2nn}
&&\left<p_t^2\right>(\vec{b},\vec{x_t},z_A,z_B)=\left<p_t^2\right>_{NN}+ \\
\ \ \ \ \ \ &&a_{gN}\rho_0^{-1}\left(T_A(\vec{x_t},-\infty,z_A)+
T_B(\vec{x_t}-\vec{b},z_B,\infty)\right),\nonumber
\end{eqnarray}
where the thickness function $T$ is defined as
$T(\vec{x_t},z_1,z_2)=\int_{z_1}^{z_2}dz\,\rho(\vec{x_t},z)$. The
constant $\sigma_{abs}$ is usually adjusted to the data from $pA$
collisions where $\Psi$'s experience only normal suppression. From
Ref.\cite{spsnormal}, we get the latest value, $\sigma_{abs}$=4.3
mb for $J/\psi$ and $\chi_c$ at SPS energy. At RHIC energy, the
d-Au experiments show that there is litte normal
suppression\cite{rhicnormal}, we take $\sigma_{abs}$ = 0 in our
calculation. The $p_t$ broadening\cite{huefner88} effect induced
by gluon-nucleon elastic scattering in the initial state is
reflected in the second term on the r.h.s of Eq. (\ref{pt2nn}).
The latest estimate of the constant $a_{gN}$ is 0.076
GeV$^2$/c$^2$fm$^{-1}$ at SPS energy\cite{na50d}, we assume that
the value is not changed at RHIC energy and the same for $J/\psi$
and $\chi_c$. The averaged transverse momentum square
$\left<p_t^2\right>_{NN}$ in nucleon-nucleon collisions is $1.15$
GeV$^2$/c$^2$\cite{na50d} at SPS and about $2.70$
GeV$^2$/c$^2$\cite{rhicnormal} at RHIC. Again we use the same
value for $J/\psi$ and $\chi_c$.

After the comparison of Eq. (\ref{ntau}) with Eq. (\ref{ntau0}),
we get the initial condition at $\tau_0$ for Eq.
(\ref{ptransport}),
\begin{equation}
f_{\Psi}(\vec{x_t},\vec{p_t},\tau_0)=(2\pi)^3 f_N
(\vec{x_t},\vec{p_t}|\vec{b})/(\tau_0\,m_t)\ .
\end{equation}
The transverse transport equation (\ref{ptransport}) can be solved
analytically with the result\cite{zhuang}
\begin{eqnarray}
\label{solution}
f_{\Psi}(\vec{x_t},\vec{p_t},\tau)&=&e^{-\int_{\tau_0}^{\tau}d\tau'
C^{\psi}_D\left(\vec{x_t}-\vec{v_t}(\tau-\tau'),\vec{p_t},\tau'\right)/m_t}\nonumber\\
&\times&f_{\Psi}(\vec{x_t}-\vec{v_t}(\tau-\tau_0),\vec{p_t},\tau_0)\
.
\end{eqnarray}
It should be noted that the leakage term, that is the second term
on the l.h.s. of Eq.\,(\ref{ptransport}), has been shown in
Ref.\cite{zhuang} to be important in the consideration of
charmonium $p_t$ distribution.

When we know the time evolution of the produced medium and the
$\Psi$ dissociation cross section in the medium, we can calculate
the number of $\Psi$ and its $p_t$ distribution at freeze-out time
$\tau_f$ with Eq. (\ref{ntau}).

\section{Numerical results }
\begin{figure}[top]
\centering
\includegraphics[totalheight=4.2in]{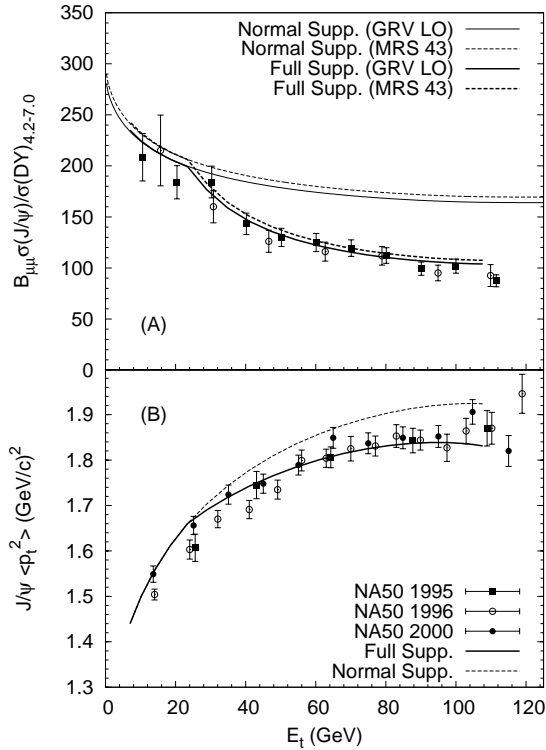}
\caption{The $J/\psi$ suppression and $\left<p_t^2\right>$ as
functions of centrality at SPS energy.} \label{sps}
\end{figure}

We first fix the parameter $s_c$ in the initialization of
hydrodynamics by fitting the $J/\psi$ suppression data from SPS.
The best fit is achieved with $s_c=31.7$ fm$^{-3}$. Fig.\ref{sps}
(A) is our calculation compared with the NA50
data\cite{spsnormal}. The thin dashed and solid lines indicate,
respectively, the theoretical calculation of normal suppression
with Drell-Yan cross section rescaled to MRS 43 and GRV LO parton
distribution function\cite{spsnormal}. The thick dashed and solid
lines are the corresponding results with anomalous suppression.
The transverse energy $E_t=0.274 N_p(b)$ GeV\cite{Blaizot2000} is
the measure of centrality with $N_p(b)$ being the average number
of participants in the collisions with impact parameter b. In very
peripheral collisions with $E_t<23$ GeV, the maximal initial
entropy density is less than the critical value $s_c$, the medium
will decouple at the initial time, and there is only normal
suppression. It is obvious that due to the missing of $E_t$
fluctuation in our calculation, the anomalous suppression starts
suddenly at $E_t=23$ GeV. With increasing centrality, the volume
and the life-time of the produced QGP will increase, and the
anomalous suppression becomes more and more important. A good fit
with the suppression data has been achieved in the whole $E_t$
range except in most central collisions where the fluctuations was
shown to be important\cite{Blaizot2000}.

After fitting the parameter $s_c$, we can calculate observables
related to the $p_t$ broadening effect. In Fig.\ref{sps} (B), we
show the $\left<p_t^2\right>$ as a function of $E_t$. The dashed
and solid lines are, respectively, $J/\psi\ \left<p_t^2\right>$
calculated without and with anomalous suppression. The later
agrees well with the experimental results from NA50\cite{na50d}.
With increasing centrality, $\left<p_t^2\right>$ increases
steadily till $E_t = 90$ GeV. After that, $\left<p_t^2\right>$
becomes saturated and finally drops slightly in most central
collisions. There are two competing mechanisms that affect
$J/\psi\ \left<p_t^2\right>$: The $J/\psi$ with large $p_t$ is
mostly produced in central collisions according to the Cronin
effect. Since the matter produced in central collisions is denser
and hotter than that in peripheral collisions, the $J/\psi$ with
large $p_t$ has more chance to be absorbed by the QGP; On the
other hand, the $J/\psi$ with large $p_t$ has more chance to
escape the anomalous suppression region, when its duration time in
the QGP is shorter than the anomalous suppression time. This is
the leakage effect initially indicated by Matsui and
Satz\cite{matsui} and studied in detail in the transport approach
in \cite{zhuang}. The former mechanism suppresses the
$\left<p_t^2\right>$, while the later enhances the
$\left<p_t^2\right>$.

\begin{figure}[top]
\centering
\includegraphics[totalheight=4.2in]{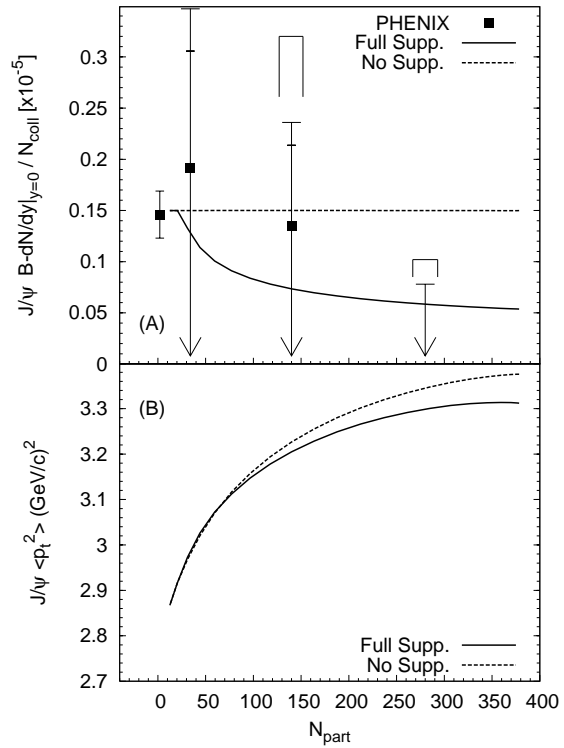}
\caption{The $J/\psi$ suppression and $\left<p_t^2\right>$ as
         functions of centrality at RHIC energy.} \label{rhic}
\end{figure}

With the fixed parameter $s_c$, we calculate now the $J/\psi$
suppression and $\left<p_t^2\right>$ at RHIC energy.
Fig.\ref{rhic} (A) shows that our result on $J/\psi$ suppression
agrees with the low statistical data from PHENIX\cite{phenix}. The
suppression at RHIC is much stronger than that at SPS, but does
not approach zero. Since there is still no transverse momentum
data from RHIC, we displayed only the theoretical results in
Fig.\ref{rhic} (B). The behavior of the $\left<p_t^2\right>$ at
RHIC is similar to the corresponding result at SPS energy.
However, it is obvious that the leakage effect prevails even in
very central collisions, and there is no clear decrease of
$\left<p_t^2\right>$, very different from the
predictions\cite{kharzeevpt, chaudhuript} calculated with $p_t$
independent dissociations and without considering the leakage
effect.

\begin{figure}[top]
\centering
\includegraphics[totalheight=2.0in]{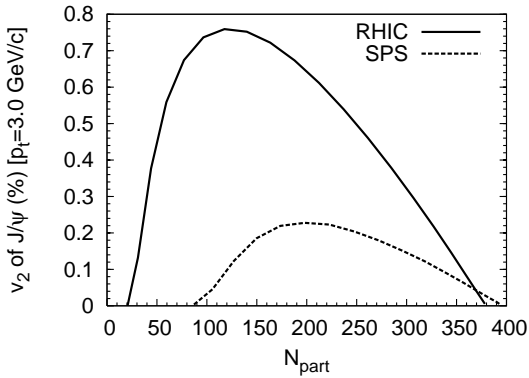}
\caption{The elliptical flow $v_2$ of $J/\psi$ as a
         function of centrality at SPS and RHIC.}\label{v2fig}
\end{figure}

Finally we consider the elliptical flow defined by
\begin{equation}
v_2 = {\left<p_x^2-p_y^2\over p_x^2 + p_y^2\right>}.
\end{equation}

In high-energy nuclear collisions, the initial spatial anisotropy
is transferred to final momentum anisotropy via partonic and/or
hadronic interactions. Interesting information about the early
stage of the hot and dense medium may be obtained from the study
of this azimuthal
anisotropy~\cite{sorge97,sorge99,ollitrault92,shuryak01,pasi01}.
The measured values of anisotropy $v_2$ of hadrons with light
quarks (u,d,s) at RHIC are positive and large compared with lower
energy results, and show a clear hadron mass dependence in the low
transverse momentum region, $p_t\leq2-3$ GeV/c
\cite{star1stv2,stark0la}. Hydrodynamic calculations have been
used to explain the observed collective expansion behavior
\cite{pasi01}. The intrinsic charm mass is large compared with the
initial temperature that might be reached at RHIC \cite{karsch04}.
Therefore, any collective motion of charmed hadrons will be a
useful tool for studying: (i) charm production mechanism in
nuclear collisions: via direct early pQCD-type creation
\cite{vogt04} or via later coalescence
\cite{coalv2,pbm03,thews04,grandchamp04}; (ii) early
thermalization of light flavors (u,d,s) in the high-energy nuclear
collisions. In the case of thermalization, if coalescence is the
dominant process for charmed hadron production, $v_2$ could be
large\cite{coalv2}, while it is predicated that the direct pQCD
produced charmed hadrons carry small or zero $v_2$. However, the
leakage effect, discussed above, will lead to finite $v_2$ for
$J/\psi$ in non-central collisions\cite{heiselberg, wang}. This
can be understood transparently: The anormalous suppression of the
$J/\psi$ depends on the length that the particle travels throught
the medium. Hence more suppression is expected in the out-of-plane
direction (y-direction) which leads to finte value of $v_2$ with
positive sign.

\begin{figure}[bottom]
\centering
\includegraphics[totalheight=2.0in]{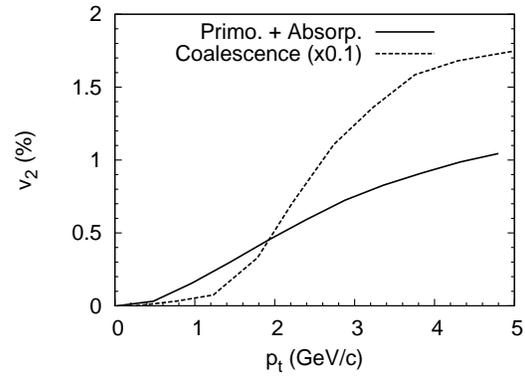}
\caption{The elliptical flow of $J/\psi$ as a function of $p_t$ at
RHIC energy. The solid line is the maximal $v_2$ with impact parameter
b=7.8 fm calculated in the frame of $J/\psi$ transport, and the dashed
line is the minimum-bias $v_2$ (scaled by a factor of 0.1) of the
coalescence model with the assumption of complete charm quark
thermalization. }
\label{v2pt}
\end{figure}

In Fig.\ref{v2fig} the solid and dashed lines are our results of
$J/\psi$ $v_2$ at fixed transverse momentum $p_t = 3$ Gev/c at SPS
and RHIC. It is clear that $v_2$ is not zero but finite in
mid-central collisions. For very central and very peripheral
collisions $v_2$ approaches to zero due to symmetric anomalous
absorption and no anomalous absorption, respectively. The values
of $v_2$ at RHIC are found to be larger than that at SPS due to
the fact that the colored medium created at RHIC has larger volume
and longer life-time which lead to the stronger leakage effect. If
the $J/\psi$'s are created at hadronization, their minimum-bias
$v_2$ has been evaluated in coalescence model with the assumption
of complete thermalization of charm quark with the medium and the
assumption of the same maximum $v_2$ (about 9\%) for charm and
light quarks\cite{coalv2}. The result is shown in Fig. \ref{v2pt}
as a function of $p_t$ and compared with our calculation at fixed
 impact
parameter b = 7.8 fm, at which the $v_2$ is found to have the
maximum. The minimum-bias $ v_2$ which is the average of $v_2$
over the impact parameter should be less than the maximum. It is
clear to see that the maximal $v_2$ calculated in the frame of
$J/\psi$ transport is less than $10\%$ of that calculated in the
approach with full charm quark thermalization \cite{coalv2}.

\section{Conclusions}
By combining the hydrodynamic equations for the QGP evolution and
the transport equation for the primordially produced $J/\psi$ in
the QGP, and considering the anomalous charmonium suppression
induced by the gluon dissociation process, we calculated the
$J/\psi$ suppression, averaged transverse momentum square, and
elliptic flow at SPS and RHIC energies. It is found that the
leakage effect reflected in the transport equation and the $p_t$
dependence of the charmonium dissociation cross sections play an
important rule in explaining the experimental $p_t$ broadening
effect at SPS.

In our transport approach, the main driving force of the elliptic
flow is the leakage effect, the $J/\psi$'s moving in the direction
of the long axis of the anisotropic medium are strongly
suppressed, but those moving in the direction of the short axis
are easy to escape the medium. The inelastic interaction with the
medium will transfer collective flow to the charmonium, but the
effect is small. Since the primordially produced $J/\psi$ is not
thermalized in our transport approach, the calculated $v_2$ is
the low limit of the $J/\psi$ elliptic flow. Any calculation with
partial or full charm quark or charmonium thermalization should be
larger than this low limit. From the comparison of theoretical
calculations in different models with the experimental data, the
elliptic flow may be a useful tool to determine which mechanism is
the dominant one of $J/\psi$ production, initial pQCD production,
coalescence at hadronization, or the mixture of the both.

The possible gain term on the right hand side of the transport
equation due to the recombination of $c$ and $\bar c$ in the
parton stage will probably enhance the charmonium production at
RHIC energy and affect $\left<p_t^2\right>$ and elliptic flow.
These effects are now under consideration.

\vspace{0.1in}
\noindent

{\bf \underline{Acknowledgments:}} We are grateful to Larry McLerran
and Bin Zhang for the helpful discussions. The work is supported by
the grants NSFC10135030, G2000077407 and the U.S. Department of Energy
under Contract No. DE-AC03-76SF00098.

\end{document}